\documentclass[sigconf]{acmart}
\usepackage{booktabs}
\usepackage{multicol}
\usepackage{color}
\usepackage{colortbl}
\usepackage{tabularx,colortbl}
\usepackage{caption}
\usepackage{multirow}
\usepackage{multicol}
\usepackage[inline]{enumitem}
\usepackage{balance}
\usepackage{subcaption}
\usepackage{url}
\usepackage{fdiaz}
\usepackage{graphics}
\usepackage{fdiaz}
\newcommand*{\tabindent}{ \hspace{3mm}}
\usepackage{xspace}

\newcommand{\plotdata}{\textsc{WikiPlot}\xspace}

\newcommand{\rqX}[1]{\textbf{RQ#1}\xspace}

\author{Jaime Arguello}
\affiliation{%
  \institution{University of North Carolina at Chapel Hill}
}
\email{jarguell@email.unc.edu}

\author{Adam Ferguson}
\affiliation{%
  \institution{Microsoft}
}
\email{adfergus@microsoft.com}

\author{Emery Fine}
\affiliation{%
  \institution{Microsoft}
}
\email{emfine@microsoft.com	}

\author{Bhaskar Mitra}
\affiliation{%
  \institution{Microsoft}
}
\email{bhaskar.mitra@microsoft.com}

\author{Hamed Zamani}
\authornote{A part of this work was done while Hamed Zamani was affiliated with Microsoft.}
\affiliation{%
  \institution{University of Massachusetts Amherst}
}
\email{zamani@cs.umass.edu}

\author{Fernando Diaz}
\authornote{Now at Google.}
\affiliation{%
  \institution{Microsoft}
}
\email{diazf@acm.org}

\copyrightyear{2021} 
\acmYear{2021} 
\setcopyright{acmcopyright}\acmConference[CHIIR '21]{Proceedings of the 2021 ACM SIGIR Conference on Human Information Interaction and Retrieval}{March 14--19, 2021}{Canberra, ACT, Australia}
\acmBooktitle{Proceedings of the 2021 ACM SIGIR Conference on Human Information Interaction and Retrieval (CHIIR '21), March 14--19, 2021, Canberra, ACT, Australia}
\acmPrice{15.00}
\acmDOI{10.1145/3406522.3446021}
\acmISBN{978-1-4503-8055-3/21/03}

\begin{document}
\title{Tip of the Tongue Known-Item Retrieval}
\subtitle{A Case Study in Movie Identification}

\begin{abstract}
While current information retrieval systems are effective for known-item retrieval where the searcher provides a precise name or identifier for the item being sought, systems tend to be much less effective for cases where the searcher is unable to express a precise name or identifier. We refer to this as \emph{tip of the tongue (TOT) known-item retrieval}, named after the cognitive state of not being able to retrieve an item from memory. Using movie search as a case study, we explore the characteristics of questions posed by searchers in TOT states in a community question answering website. We analyze how searchers express their information needs during TOT states in the movie domain. Specifically, what information do searchers remember about the item being sought and how do they convey this information? Our results suggest that searchers use a combination of information about: (1) the \emph{content} of the item sought, (2) the \emph{context} in which they previously engaged with the item, and (3) previous attempts to find the item using other resources (e.g., search engines). Additionally, searchers convey information by sometimes expressing uncertainty (i.e., hedging), opinions, emotions, and by performing relative (vs. absolute) comparisons with attributes of the item. As a result of our analysis, we believe that searchers in TOT states may require specialized query understanding methods or document representations. Finally, our preliminary retrieval experiments show the impact of each information type presented in information requests on retrieval performance.
\end{abstract}
\maketitle

\section{Introduction}\label{sec:introduction}

Known-item retrieval refers to a broad class of scenarios where a searcher's information need is for a single, specific document known to exist in the corpus \cite{Lee2006}. Searchers may seek an item they have seen before \cite{Dumais2003} or one they believe exists \cite{broder:web-search-taxonomy}. During known-item retrieval, searchers may express their need using a unique identifier (e.g., title), bibliographic information (e.g., author or genre), or content cues (e.g., keywords). Information retrieval systems, for example web search engines, can leverage previously issued queries and engagements to improve known-item search, although such techniques are less effective for unpopular or new documents with less behavioral data \cite{downey:headstails}.

In this paper, we investigate known-item retrieval scenarios where a searcher is looking for a previously seen item but is \emph{not} able to express precise or even reliable information about the item.  Imprecision may result from a long delay between the information need and when the searcher most recently engaged with the item (i.e., long-term memory degradation) or from the lack of a universally adopted content description language (e.g., searching for a song based on its drum beat or a book based on its narrative structure).  While similar to re-finding tasks, the emphasis on identification---as opposed to navigation---makes these information needs acute, one-off episodes---as opposed to repeated requests.  

We refer to a searcher in this situation as being in a `tip of the tongue' (TOT) state, due to its similarity to the cognitive state of not being able to retrieve an item from memory \cite{brown:tot}.
We adopt the following definition of a TOT information need,
\begin{quote}
\textit{an item identification task where the searcher has previously experienced or consumed the item but cannot recall a reliable identifier.}
\end{quote}
\citet{Elsweiler2007} found that searchers specifically in TOT states tend to be prone to higher levels of frustration compared to other memory lapse states.

In recent years, several community question-answering (CQA) sites have emerged to support searchers with TOT information needs.\footnote{\url{https://www.reddit.com/r/tipofmytongue}, \url{https://irememberthismovie.com}, \url{https://www.watzatsong.com/en}, \url{ https://www.goodreads.com/group/show/185-what-s-the-name-of-that-book}, \url{ https://scifi.stackexchange.com/questions/tagged/story-identification}}  These sites are tailored for searchers who want to find a specific known item (e.g., a movie, book, song, artist, video game, childhood toy) but do not remember its name or other unique metadata.  Searchers pose questions, composed of a title and description, while community members ask clarification questions and suggest answers.  A thread is typically closed (or ``solved'') when the questioner indicates that the correct answer has been provided.  Importantly, one might view such CQA sites as serving people with TOT information needs that could not be resolved using other resources, \emph{including search systems}~\cite{liu2012searchfails}.  To paraphrase a blogger's comment about \verb|watzatsong.com|, ``when computers fail, seek the help of humans.''\footnote{\url{https://www.labnol.org/internet/find-name-of-songs/12316/}} 

In order to study TOT information needs, we analyzed questions (i.e., information \textit{requests}) posted to the `I Remember This Movie' TOT CQA site.\footnote{\url{https://irememberthismovie.com/}} We decided to use movie identification as an initial case study because the movie domain is associated with rich auxiliary data available online, including metadata and long-form plot descriptions. 

Our research in this paper focuses on two main research questions.  First, we are interested in understanding TOT search requests,
\begin{quote}
\rqX{1}: How does a searcher in a TOT state express their need in the search request?
\end{quote}
To answer this question, we investigated the characteristics of TOT information requests.  Our goal is to understand the types of phenomena present in TOT requests. Insights gained as part of \rqX{1} help us answer two basic questions:
\begin{enumerate*}
\item What do searchers in TOT states remember? and
\item How do they convey this information?  
\end{enumerate*}
To address \rqX{1}, we conducted an extensive qualitative analysis of 1,000 requests posted to `I Remember This Movie'.  Our qualitative coding scheme (applied at the sentence-level) was developed in a bottom-up fashion.  After analyzing a subset of the data, we developed qualitative codes along four dimensions: 
\begin{enumerate*}
\item the \emph{target} of the sentence (i.e., what is the sentence about?), 
\item the presence of opinions or emotional expressions, 
\item expressions of uncertainty (i.e., hedging), and 
\item the presence of relative comparisons (versus descriptions in absolute terms).  
\end{enumerate*}
In terms of the \emph{target}, sentences described characteristics of the movie itself, the \emph{context} in which the searcher watched the movie, previous (failed) attempts to find the movie, or neither (i.e., social nicety).

We were also interested in understanding the performance of automatic retrieval for TOT needs,
\begin{quote}
\rqX{2}: How well does a conventional retrieval system satisfy TOT requests?
\end{quote}
To answer this question, we gathered a corpus of movie plot descriptions and evaluated the effectiveness of TOT requests when issued as queries to a standard information retrieval system.  We studied the relationship between retrieval performance and the presence/absence of codes developed as part of \rqX{1}.  For example, does performance improve if we ignore sentences that describe the \emph{context} in which the searcher watched the movie? Or, does performance improve if we ignore sentences that convey uncertainty?

Our results provide insights about the behaviors of searchers in TOT states and the effectiveness of existing IR systems in responding to TOT information requests. Our results suggest that searchers in TOT states tend to leverage memories about the movie itself (e.g., scenes, characters, locations) and the contexts in which they viewed the movie (e.g., time, place, and even concurrent external world events). The presence of contextual information in TOT requests is consistent with previous results from the personal information management (PIM) literature~\cite{Dumais2003,Elsweiler2007,Elsweiler2005,Hwang2017}. While searchers may not recall the exact keywords in an email, they may recall things that were happening when the email was received. Additionally, our results suggest that searchers use a variety of tactics that are not supported by conventional search systems, including multi-hop reasoning (e.g., comparisons to other movies), self-reflective descriptions of previous search attempts, and expressions of uncertainty.  In terms of retrieval performance, our results suggest that current IR systems can successfully leverage descriptions of the movie, but not descriptions of the context. Furthermore, they are surprisingly robust to expressions of uncertainty. We discuss opportunities for future research to develop algorithmic solutions to resolve these types of information needs.

\section{Related Work}\label{sec:related_work}

Our research builds on three areas of prior work. First, our work builds on psychology research on long-term memory, asking what people remember and how they convey memories. Similarly, our work builds on prior information retrieval research aimed to support information \emph{re-finding} in situations where a searcher has lapses in memory. Much of this research has been done in the context of \emph{personal information management} (PIM). Finally, prior IR research has also studied how information requests aimed at human intermediaries differ from requests aimed at a search system.

\paragraph{\textbf{Long-term Memory}} Long-term memory plays an important role in information re-finding. It is generally accepted that human memory is transient. Research in psychology has studied how different factors contribute to lapses in long-term memory. For example, research clearly shows that long-term memory degrades with time (i.e., decay theory~\cite{Rubin1995}) and by an individual engaging with new tasks and information objects (i.e., interference theory~\cite{Bower1994}). Additionally, memory degrades quicker when an individual did not explicitly aim to remember the item in question (i.e., poor encoding~\cite{Elsweiler2007}). In the context of movies, recall may degrade with time and by a searcher engaging with other (perhaps similar) movies. Additionally, gaps in memory may occur simply because the searcher did not aim to remember the movie in the long term.

Psychology research has studied, not only \emph{how much} is remembered, but also \emph{what} is remembered. For example, research shows that people tend to forget precise details and remember high-level characteristics or ``gists''~\cite{Rubin1977}. In the context of movies, a searcher may forget the name of a character, but remember their personality. Additionally, models of long-term memory distinguish between \emph{declarative} memory (i.e., remembering bits of information) versus \emph{procedural} memory (i.e., remembering skills). Furthermore, declarative memory is sub-divided into \emph{semantic} memory (i.e., memory about inherent characteristics of an information item) versus \emph{episodic memory} (i.e., memory about previous engagements with the item)~\cite{Rugg2000,Tulving1993}. Episodic memory can be viewed as ``autobiographical'' and deals with subjective experiences.

Previous studies about \textit{how} people recall have demonstrated consistent strategies used by individuals. For example, when presented with the photograph of a famous person, individuals leverage information about the person (e.g. profession, places) in order to recall their name \cite{yarmey:tot-faces}.  This is an example of the `tip of the tongue' phenomenon, `a state in which one cannot quite recall a familiar word but can recall words of similar form and meaning' \cite{brown:tot,schwartz:tot-survey}. Compared to other information seeking tasks, both the information retrieval \cite{Elsweiler2007} and cognitive psychology \cite{metcalfe:tot-curiosity} literature indicate that searchers in TOT states exhibit more frustration at not knowing the answer and more satisfaction when the answer is revealed.

Interestingly, psychology research has used movies to study long-term memory in a controlled laboratory setting. As noted in \citet[p457]{furman2007they}, using movies to study long-term memory is appealing because they can simulate ``aspects of real-life experiences by fusing multimodal perception with emotional and cognitive overtones.'' \citet{furman2007they} conducted a study in which participants watched a 30-minute movie and completed tests to measure long-term memory at different times, ranging from 3 hours to 9 months after watching the movie. As one might expect, test performance \emph{and} self-reported confidence in the test answers degraded over time. Interestingly, however, performance degraded differently for test questions about different aspects of the movie. For example, performance degraded quicker for questions that asked about specific details (e.g., verbatim quotes) than questions that asked about themes and scenes involving social interactions.

\paragraph{\textbf{Memory and Personal Information Management (PIM)}} PIM research studies how people manage and access their personal information (e.g., files, emails, photos, etc.). Memory plays an important role in PIM. Much PIM research has studied the importance of episodic memory (versus only semantic memory) during search and re-finding. In other words, PIM studies suggest that systems should support searching and re-finding using contextual cues. For example, a searcher may not remember the contents of an email in order to create an effective keyword query but may remember the day the email was received or events that happened that day.

\citet{Dumais2003} evaluated the ``Stuff I've Seen'' system and found that ``last modified date'' was the most widely used contextual cue for filtering search results. \citet{Elsweiler2007} conducted a diary study focused on participants' everyday memory problems and strategies used to overcome these. Results confirmed the importance of episodic memory to support information re-finding---participants often forgot details about the item itself, but remembered contextual details about when the item was last used (e.g., the task they were trying to accomplish when they last engaged with an item). \citet{Elsweiler2005} evaluated a search system for managing, tagging, and accessing personal photographs. The system allowed searchers to search (and re-find) based on episodic memories. Results found that participants often searched based on \emph{multiple} contextual cues (e.g., time, place, event-type, etc.). \citet{Hwang2017} evaluated a system to tag bookmarked pages with contextual information. For example, participants could tag bookmarks based on time, location, and the task the participant was working on when the bookmark was made. Results found that contextual cues were effective in helping participants re-find bookmarks. Additionally, the task associated with a bookmark was the most recalled contextual cue.

In the context of life-logging, prior work also suggests that episodic memories (combined with semantic memories) can help searchers find information within their own \emph{human digital memory} (HDM) repositories~\cite{Fuller2008}. Similarly, \citet{Kelly2008} compared the effectiveness of queries containing only content information (i.e., leveraging semantic memory) versus queries containing both content and context information (i.e., leveraging semantic and episodic memory). Queries combining content-related and contextual cues outperformed content-only queries.

\paragraph{\textbf{Information Requests aimed to Humans}} \citet{Wilson1999} proposed that information needs gradually evolve over four stages. First, a \emph{visceral} need is one that cannot be expressed in words---there is a vague sense of unease that cannot be explained. Second, a \emph{conscious} need is one that remains ambiguous (i.e., the searcher does not know what information is needed) but could be potentially resolved by talking to others. Third, a \emph{formalized} need is one that can be communicated to others, but perhaps not to a search system. Finally, a \emph{compromised} need is one that can be formulated using a specific interface or query language (e.g., by choosing specific keywords). In this paper, we study TOT information requests posed to human intermediaries. One might argue that these are cases where a searcher does not have the requisite knowledge to transition from a formalized state to an effective compromised state. To build better information retrieval systems, it is important to study not only \emph{compromised} information needs (e.g., how people formulate keyword queries), but also \emph{formalized} information needs (e.g., how people convey information needs to a human intermediary).

\citet{Arguello2018} conducted a large-scale user study that compared information requests (for the same information needs) aimed to a search system versus a human intermediary. Additionally, they considered information needs associated with specific types of \emph{extra-topical} relevance criteria (e.g., temporal-, geographical-, complexity-related criteria). Through qualitative analysis, the authors compared the search strategies adopted by searchers when conveying extra-topical relevance criteria during requests aimed at a human intermediary versus a search system. In the human intermediary condition, participants reported less difficulty producing their requests but adopted search strategies that deteriorated retrieval performance when the request was issued to a web search engine. For example, in the human intermediary condition, participants were more likely to convey what they \emph{did not} want.

\citet{KatoSIGIR2014,KatoWWW2014} also studied how people formulate requests when the information need has a specific extra-topical dimension (e.g., domain knowledge). In this case, requests were aimed to a search system. Results found that participants often ignored the extra-topical dimension in their queries or used ``indirect'' strategies that work well with current search systems (e.g., using the query-term `Wikipedia' to get results for a domain novice). Queries performed poorly when they explicitly mentioned the extra-topical dimension (e.g., `simple explanation...').

More closely related to our research, \citet{Hagen2015} curated a corpus of known-item questions posed to the Yahoo!~Answers Q\&A site. Most questions aimed at (re-)finding a website.  However, some questions aimed at (re-)finding a previously experienced item (e.g., movie, book, song, band/musician, etc.).  Interestingly, a qualitative analysis found that 240 questions (out of 2,755) contained so-called ``false memories'', in which the asker provided incorrect information (e.g., mentioning the wrong actor when trying to recollect a movie). In a recent paper, \citet{Jorgensen2020} report on qualitative analysis of TOT requests posed to the `Tip of my Joystick' Reddit community, aimed at helping people re-find previously played video games. The coding scheme developed focused on similar phenomena as ours---visual characteristics (e.g., characters), audio characteristics (e.g., soundtrack), metadata characteristics (e.g., release date), comparisons with other video games, and characteristics of the \emph{context} in which the asker previously engaged with the video game.  The authors discuss controlled vocabularies necessary to support TOT requests for video games.

\section{Movie Identification}\label{sec:data}

Query-based search is an important component of movie streaming services such as Netflix and YouTube. However, existing systems tend to rely on metadata (e.g., title, actor, director, genre, time period) as opposed to plot or scene descriptions~\cite{liu:conversational-movie-search,lamkhede:netflix-search,hosey:search-mindsets}. In our case, we focus on TOT information requests in which the searcher does not remember metadata information or the intent is not easily expressible as a keyword query.

Perhaps due to the failure of current search systems to support searchers in TOT states, several community question answering sites have gained popularity on the web. These sites allow searchers to pose a question, composed of two parts: a title for the question and a longer, free text description of the item. In response to a question, a conversational thread consists of community members asking clarifying questions and suggesting answers while searchers respond to questions and indicate when an answer is correct.

We collected questions posted to `I Remember This Movie', a community question-answering site specifically designed for individuals seeking to identify movies that they have seen but whose title they cannot recall.  Because we could not confirm that all questions in fact referred to real movies, we restricted our collection to questions with correct answers, as indicated by the searcher.  This resulted in a set of 2,072 questions posted between 2013 and 2018. Of these, 762 correct answers included links to IMDb pages, which we recorded for further analysis.

\section{Analysing TOT Requests}\label{sec:annotation}
In order to better understand TOT requests and address \rqX{1}, we performed a qualitative analysis of the requests from the dataset described in Section \ref{sec:data}.  Our analysis operated at the granularity of sentences, to allow easier annotation and interpretability of results. We developed a qualitative coding scheme to be applied at the sentence level and then analysed these annotations on our corpus.  

\subsection{Coding Scheme}\label{subsec:requests:methods}
We segmented TOT requests into sentences using the Stanford NLP toolkit\footnote{\url{https://nlp.stanford.edu/software/}} and applied qualitative codes to all sentences from a randomly sampled set of 1,000 requests.

Our coding scheme was developed in a top-down and bottom-up fashion. First, after looking at the data, four of the authors agreed on \emph{seven} broad categories for codes: \textit{movie}, \textit{context}, \textit{previous search}, \textit{social}, \textit{uncertainty}, \textit{opinion/emotion}, and \textit{relative comparison}. Four of these categories (\textit{movie}, \textit{context}, \textit{previous search}, and \textit{social}) focus on the topic of the sentence. The sentence may describe the movie, the context in which the movie was seen, a previous attempt to find the movie, or merely be a social nicety.  The remaining three categories (\textit{uncertainty}, \textit{opinion/emotion}, and \textit{relative comparison}) relate to interesting phenomena that we noticed and wanted to capture in our coding scheme. We noticed sentences that express uncertainty about the information being conveyed, express opinions about an aspect of the movie, describe emotional states, and provide descriptions in relative terms by drawing comparisons.

After identifying these seven broad categories, codes were developed in a bottom-up fashion by two of the authors. Codes were developed in three phases. During the first phase, two of the authors independently developed their own codes and then met to discuss their codes and definitions. At this point, both authors developed a preliminary coding scheme (i.e., codes and definitions) using the union of their individual coding schemes. During the second phase, both authors independently coded 50 randomly sampled sentences. After measuring inter-annotator agreement, codes with low agreement (i.e., Cohen's $\kappa\le .20$) were redefined, dropped, or combined with other codes. Finally, during the third phase, both of the authors independently coded 758 randomly sampled sentences. At this point, inter-annotator agreement for all codes was satisfactory. Our final coding scheme (described below) resulted in 34 codes. Of these, 2 codes (6\%) had a Cohen's $\kappa$ agreement at the level of `fair', 5 (15\%) at the level of `moderate', 16 (47\%) at the level of `substantial', and 9 (26\%) at the level of `almost perfect'~\cite{Landis77}. All codes were designed to not be mutually exclusive, meaning that sentences could be assigned zero, one, or more codes. 

After developing and testing the reliability of our coding scheme, both coders coded 1,000 TOT information requests (500 each) comprised of 8,030 total sentences. Our coding scheme was comprehensive in terms of capturing a wide range of phenomena. As it turns out, all sentences coded were assigned at least one of our 34 codes.

\subsection{Results}\label{subsec:requests:results}
Our codes are described in Table~\ref{tab:annotation-results-movies} (content codes), Table~\ref{tab:annotation-results-context} (context codes), and Table~\ref{tab:annotation-results-others} (other codes). In each table, the first column contains the code label, the second column provides a definition, the third column illustrates an example sentence that was assigned the code, the fourth column shows the relative frequency of the code (i.e., the percentage of sentences associated with the code), and the fifth column shows the Cohen's $\kappa$ agreement for the code. As previously mentioned, to test the reliability of our coding scheme, two of the authors independently coded the same set of 758 randomly sampled sentences. To measure inter-coder agreement, we computed the Cohen's $\kappa$ agreement for each code independently. Cohen's $\kappa$ ranges from -1 to +1 and measures agreement after correcting for agreement due to random chance. A value of -1 signals perfect disagreement, +1 signals perfect agreement, and 0 signals agreement no better than random. The relative frequencies (fourth column) in Tables~\ref{tab:annotation-results-movies}--\ref{tab:annotation-results-others} do not sum to 100\% because codes were not mutually exclusive.

\paragraph{\textbf{Movie}} As expected, many sentences described characteristics of the movie itself. As shown in Table~\ref{tab:annotation-results-movies}, these sentences include descriptions of the characters in the movie, a scene, a physical object, the movie's category/genre/tone, and the overall plot. This high-level dimension focuses on \emph{semantic} memories---recollections about the movie itself. 

\paragraph{\textbf{Context}} We also noticed that some sentences describe the context in which the searcher previously engaged with the movie. As shown in Table~\ref{tab:annotation-results-context}, these included references to when and where the searcher watched the movie, the medium (e.g., TV, movie theatre, etc.), who they watched it with, and even world events that were happening around the time period they watched the movie. This high-level dimension focuses on \emph{episodic} memories---recollections about the context in which the searcher previously engaged with the movie. As noted in Section~\ref{sec:related_work}, episodic memories play an important role during information re-finding in the context of personal information management~\cite{Dumais2003,Elsweiler2007,Elsweiler2005,Hwang2017}. For example, \citet{Dumais2003} found that one might not remember the content of an email, but may remember when they received it or what was happening when they received it. The annotations based on our context codes suggest that episodic memories also play a role in TOT information needs.

The remaining codes are described in Table~\ref{tab:annotation-results-others}. 

\paragraph{\textbf{Previous Search}} The previous search dimension focuses on references to failed attempts to identify the movie title. These include references to previous information sources consulted (e.g., search engines, websites, people), as well as descriptions of search strategies that were unsuccessful (e.g., searching through an artist's filmography).

\paragraph{\textbf{Social}} This category relates to sentences containing social niceties (e.g., please and thank you) \cite{dnz:politeness}.  Sentences were labeled at a high rate (10.77\%) compared to other groups of annotations.

\paragraph{\textbf{Uncertainty}} Interestingly, we also noticed that searchers often used linguistic markers of uncertainty (e.g., ``I vaguely remember that...'').  The high rate of this annotation (35.37\%) suggests that searchers in TOT states are self-aware when providing incomplete or unreliable information due to long-term memory degradation.

\paragraph{\textbf{Opinion/Emotion.}} The opinion/emotion dimension focuses on the presence/absence of opinionated statements and references to emotional states. Here, we refer to opinions as judgements, critiques, or evaluations about some aspect of the movie. Conversely, we refer to emotional expressions as references to emotional states experienced by the searcher while watching the movie.

\paragraph{\textbf{Relative Comparison}} This dimension relates to whether the sentence conveys information in relative versus absolute terms. An absolute statement is one that conveys information without drawing comparisons that require background knowledge or additional informational in order to extract its full meaning. For example, ``The man character is a blond, handsome man.'' is an absolute statement. Conversely, a relative statement is one that draws one or more comparisons that require additional information in order to extract the meaning of the statement. For example, ``The main character looks like Brad Pitt.'' is a relative statement. Extracting the meaning of this sentence requires resolving Brad Pitt's physical features. We view relative comparisons as statements that require some degree of inference using background knowledge. In other words, relative comparisons require multi-hop reasoning.

\begin{table*}
 \centering
 \caption{\textit{Movie Annotation}. Codes related to characteristics of the movie.}
 \label{tab:annotation-results-movies}
 {\footnotesize
	\begin{tabular}{lp{2in}p{2.9in}rr}
 \toprule
code & definition & examples & frequency & $\kappa$ \\
 \hline
Character & Describes a character. & The main protagonist is a 20-something girl with short hair, which is either blonde or brunette. & 51.21\% & 0.766 \\
Scene & Describes a scene. & Finally the real boyfriend appeared at the final scene in a cabin near some lake or sea and they try to kill each other. & 36.53\% & 0.755 \\
Object & Describes a tangible object in a scene. & They're in the car and they almost crash into this beast. & 26.72\% & 0.750 \\
Category & Describes the movie category (e.g., movie, tv movie, miniseries, etc.). & Live-action possibly made-for-TV. & 25.07\% & 0.536 \\
Location type & Describes a scene's location type. & The movie starts out with an American family who are staying in some Eastern European Castle with their young son. & 18.36\% & 0.698 \\
Plot summary & Describes the overall plot or premise. & This movie is about a young girl who marries early and has a baby boy. & 10.82\% & 0.637 \\
Release date & Describes timeframe of movie release. & I remember this horror movie from late 70s early 80s. & 5.43\% & 0.854 \\
Genre/tone & Describes genre or tone. & I think it was a romantic comedy of sorts. & 5.39\% & 0.782 \\
Visual style & Describes visual style (e.g., black and white, colour, CGI animation, etc.). & It was in English not subtitled and in colour. & 4.73\% & 0.821 \\
Language & Describes the language spoken. & The dialog in the movie was in Spanish. & 2.89\% & 0.955 \\
Regional Origin & Describes movie's region of origin. & I think it was an European movie and not in English. & 1.72\% & 0.933 \\
Specific location & Describes a scene's specific location. & I believe they were traveling to Louisiana to pick up a friend's body for a funeral. & 1.58\% & 0.598 \\
Quote/dialogue & Describes a quote from the movie. & The wife yells something along the lines of either `look what you did' or `look what you did to my husband'. & 1.54\% & 0.766 \\
Real person & Describes real person associated with movie (directly or indirectly). & The woman looked like Annie Clark . & 1.21\% & 0.864 \\
Camera angle & Describes camera action. & The jumping between scenes was also very strange. & 0.95\% & 0.663 \\
Singular timeframe & Describes timeframe. & I think it was made in either the 70s or 80s but the movie is set in the 20s or 30s. & 0.71\% & 0.832 \\
Multiple timeframe & Describes the passage of time in the movie. & Decades later the house that is above the tunnel is believed to be haunted. & 0.63\% & 0.799 \\
Fictional person & Describes fictional person associated with movie (directly or indirectly). & It was a scene with two adventures in a scene like Indiana Jones trapped captured by enemy forces. & 0.62\% & 0.712 \\
Actor nationality & Describes nationality or ethnicity associated with actor/actress. & She is a regular height woman also Caucasian slim and with red hair. & 0.54\% & 0.499 \\
Target audience & Describes movie's target audience. & Gadget packed action movie for Kids? & 0.49\% & 1.000 \\
Compares music & Describes movie's soundtrack. & I remember there was lots of nice electronic music but what was the title of the movie? & 0.32\% & 0.888 \\
Specific music & Describes specific song in the movie. & The mother makes her living from singing in small joints at some point she sings a version of "Looking for the Heart of Saturday Night". & 0.15\% & 0.666 
 \\
 \bottomrule
 \end{tabular}}
\end{table*}

\begin{table*}
 \centering
 \caption{\textit{Context Annotation}. Codes related to characteristics of the context surrounding the searcher's previous engagement with the movie.}
 \label{tab:annotation-results-context}
 {\footnotesize
	\begin{tabular}{lp{2in}p{2.9in}rr}
 \toprule
code & definition & examples & frequency & $\kappa$ \\
 \hline
Temporal context & Describes when the movie was seen, either in absolute terms (e.g., around 2008) or relative terms (e.g., when I was a kid). & I rented this film in the early 2000's. & 8.58\% & 0.783 \\
Physical medium & References the physical medium associated with watching the movie (e.g., TV, theatre, VHS, etc.) & I remember it was like in the 2000's and it was on the tele. & 5.42\% & 0.855 \\
Cross media & Describes exposure to movie through different media (e.g., trailer, DVD cover, poster, etc.) & One of its posters shows a man waving a sheet of white cloth. & 1.06\% & 0.542 \\
Contextual witness & Describes other people involved in the movie-watching experience. & I remember it was on television and I was so young my parents made me turn it off. & 0.76\% & 0.621 \\
Physical location & Describes physical location where movie was watch. & I watched this movie in my film class a couple years ago. & 0.72\% & 0.621 \\
Concurrent events & Describes events relevant to time period when movie was watched. & I've seen it around 2006 ( I know cause I watched it alongside Hard Candy). & 0.14\% & 1.000
 \\
 \bottomrule
 \end{tabular}}
\end{table*}

\begin{table*}
 \centering
 \caption{\textit{Other Annotations}. Codes associated with previous search attempts, social niceties, uncertainty, opinions, emotions, and relative comparisons.}
 \label{tab:annotation-results-others}
 {\footnotesize
	\begin{tabular}{lp{2in}p{2.9in}rr}
 \toprule
code & definition & examples & frequency & $\kappa$ \\\hline
Previous search & Describes a previous attempt to find the movie title. & I tried to find it using google, searched a number of databases with sci-fi movies from 1960s-1990s with no success. & 1.48\% & 0.811\\\hline
Social & Communicates a social nicety & If you could help at all I'd really appreciate it! & 10.77\% & 0.735 \\ \hline
Uncertainty & Conveys uncertainty about information described. & It was a foreign film I think either French or German, but I could be wrong. & 35.37\% & 0.512 \\\hline
Opinion & Conveys an opinion or judgement about some aspect of the movie. & Its pretty confusing all the way to the end when there's only one surviving woman and then she is sat in the same room with this monster. & 2.09\% & 0.341 \\
Emotion & Conveys how the movie made the viewer feel. & It was the first movie that kept me awake at night. & 0.46\% & 0.283 \\\hline
Relative comparison & Describes a characteristic of the movie in relative (vs. absolute) terms. & One of the detectives is young laid back kinda like Kevin Bacon or Gary Sinise but looking through their filmography I could not find the movie. & 3.01\% & 0.701
 \\
 \bottomrule
 \end{tabular}}
\end{table*}

\subsection{Discussion}\label{subsec:requests:discussion}

Our qualitative analysis of TOT requests reveal several important trends. These trends provide insights about: (1) the things people remember and (2) the things people decide to convey when attempting to resolve their TOT information needs.

\paragraph{\textbf{What people remember}} First and foremost, our analysis reveals that people remember characteristics of the movie (e.g., a scene, character, object) as well as characteristics of the context in which the movie was seen (e.g., time, place, physical medium, external events). In other words, it appears that searchers rely on both semantic \emph{and} episodic memories when attempting to resolve a TOT information need.

Second, based on our coding scheme, searchers conveyed visual memories more than auditory memories. Our most frequent codes (>18\%) involved
visual memories (e.g., character, scene, object, location type). Codes associated with auditory memories (e.g., quotes, compares music, specific music) were much less frequent (< 2\%). We see at least two possible explanations for this trend. One possibility is that visual memories are easier to communicate than auditory memories. In other words, perhaps searchers had plenty of auditory memories, but they decided to omit them in their TOT queries. Alternatively, it is possible that searchers had more visual memories than auditory ones. This explanation would be consistent with prior research that has found that visual memory is more robust (i.e., long-lasting) than auditory memory~\cite{Brady2008,Cohen2009}.

Third, most of our frequent codes are related to things that exist in the physical world and can be perceived by the senses (e.g., character, scene, object, location type). Only one of our codes (i.e., tone) relates to an abstract characteristic of movies (e.g., dark, scare, fantasy). This trend also seems consistent with prior work that has found that memories of concrete characteristics are more robust than abstract ones~\cite{Marcel1974}.

\paragraph{\textbf{What people say}} In terms of what searchers communicate, our results suggest four important trends. First, searchers convey information that may be useful for a human intermediary (with domain and world knowledge), but potentially problematic for existing search systems that rely (partly or entirely) on keyword matching. In particular, about 10\% of sentences contained descriptions of the context in which the searcher previously watched the movie. These references (based on episodic memories) may be helpful for an intelligent human searcher, but require some degree of inference using real-world knowledge. Table~\ref{tab:annotation-results-context} provides some examples. For instance, mentioning that the movie was seen in a ``film class'' (\emph{physical location}) implies that the movie is probably artistic or noteworthy; mentioning that ``I was so young my parents made me turn it off'' (\emph{contextual witness}) implies that the movie is not appropriate for children; and mentioning that ``I watched it alongside Hard Candy'' (\emph{concurrent events}) implies that the movie came out around 2005.

Secondly, it is interesting that searchers mentioned previous (failed) attempts to find the movie. To gain more insight about these references, we examined the codes with the highest degree of co-occurrence with our \emph{previous search} code. Table~\ref{tab:pmi-search} shows all codes with a positive point-wise mutual information (PMI) score with \emph{previous search}. The first column provides the co-occurring code, the second column provides an explanation, the third column provides the PMI score, and the fourth column provides an example sentence. Four codes had positive PMI values: \emph{real person}, \emph{relative comparison}, \emph{genre/tone}, \emph{release date}. Based on the examples, searchers contributed potentially useful \emph{negative} evidence that may help someone identify the movie. The examples point to failed attempts to find the movie by searching for all movies from a certain actor/actress, all episodes of a specific series, and all movies from a given genre and release date. This information might help someone define the search space (i.e., rule out particular alternatives). This trend suggests that systems to support TOT information needs may benefit from accommodating negative feedback.

\begin{table*}
 \centering
 \caption{Codes with highest pointwise mutual information (PMI) with `Previous search'.}
 \label{tab:pmi-search}
 {\footnotesize
	\begin{tabular}{p{.8in}p{2in}p{.25in}p{3in}}
 \toprule
Co-occurring Code & Explanation & PMI & Example \\\hline
Real person & Searched by potential actor/actress. & 3.564 & For some reason I remember it as Julia Stiles but I looked at IMDb and nothing on her filmology page rings any bells. \\
Relative comparison & Searched by comparing with similar/related items. & 2.617 & My grandparents were fairly "proper" people so I expected this to be an episode of Masterpiece Mystery or Poirot but I can't find it. \\
Genre/tone & Searched by genre/tone. & 0.64 & I've gone through countless lists like "50 weird SciFi movies from the 80's" and still nothing. \\
Release date & Searched by release date. & 0.308 & I tried manually browsing wikipedia page of scifi movies from 70ies up to now but I can't seem to find it. 
 \\
 \bottomrule
 \end{tabular}}
\end{table*}

Third, relative comparisons (versus absolute statements) were found in ~3\% of all sentences. This result suggests that it is often easier for someone to draw a comparison (e.g., ``looks like Kevin Bacon'') than to describe someone or something in absolute terms. Table~\ref{tab:pmi-relative} shows all codes with a positive PMI score with respect to \emph{relative comparison}. Interestingly, common relative comparisons included comparisons with fictional characters (e.g., ``tarzan''), real people (e.g., Kirsten Dunst), time periods (e.g., Victorian-esque), genres/tones (e.g., ``let's hunt humans for fun type movie''), regional origins (e.g., ``American-style''), and specific locations (e.g., ``Grand Canyon-esque''). This trend suggests that systems to support TOT information needs may need to accommodate comparisons between people and other movie attributes. Prior work has found that information retrieval systems perform poorly on queries containing relative (versus absolute) statements~\cite{Arguello2018}.

\begin{table*}
 \centering
 \caption{Codes with highest pointwise mutual information (PMI) with `Relative comparison'.}
 \label{tab:pmi-relative}
 {\footnotesize
	\begin{tabular}{p{.8in}p{2in}p{.25in}p{3in}}
 \toprule
Co-occurring Code & Explanation & PMI & Example \\\hline
Fictional person & Comparisons with a fictional character. & 4.408 & He looked similar to Tarzan, but he wore pants and had some kind of weapon strapped across his back. \\
Real person & Comparisons with other actors/actresses or comparisons with the artistic styles of other writers/directors. & 3.976 & For some reason, I swear Kirsten Dunst was in this movie and keep thinking it has something to do with The Virgin Suicides but it is not that movie and I can not find it when searching Kirsten Dunst. \\
Previous search & Comparisons with other movies or artists in the context of a prior search attempt. & 2.617 & This'll be an easy one for you guys I'm sure, but Googling just brings up "The Craft" and "Slugs". \\
Opinion & Opinionated comparisons. & 2.614 & And it has that 70's horrible sound quality, especially when someones screaming or he's doing his creepy laugh. \\
Singular timeframe & Comparisons with temporal periods. & 2.027 & It was set in a Victorian-esque setting with horse drawn carriages. \\
Genre/tone & Comparisons with styles of other films. & 1.720 & It was definitely a "Let's hunt humans for fun" type movie. \\
Cross media & Comparisons with media associated with other movies. & 1.450 & The cover seemed almost like a National Lampoon cover. \\
Regional origin & Comparisons with movies from specific origin. & 1.114 & Also, it was translated to Turkish but the movie itself looked very American-style children's movie. \\
Specific location & Comparisons between a scene location and a real one. & 1.064 & The river looked "Grand Canyon-esque" we find out one of the girls mom gets beat up really bad by her husband and at the end he gets arrested.
 \\
 \bottomrule
 \end{tabular}}
\end{table*}

Finally, it is noteworthy that expressions of uncertainty were so common. Roughly 35\% of all sentences contained expressions of uncertainty. In linguistics, hedging allows speakers and writers to signal caution or probability versus full certainty. Again, Table~\ref{tab:pmi-uncertainty} shows all codes with a positive PMI score with respect to \emph{uncertainty}. Searchers expressed uncertainty about the movie's release date, the regional origin, an actor/actress in the movie, the lyrics of a song, the movie's timeframe, the target audience, a specific location, a specific musical piece, and even the temporal context when the movie was watched. This trend suggest that systems to support TOT information may need to deal with (un-)certainty. While information retrieval systems have not been designed to model uncertainty of a searcher's input, hedging has been incorporated in other types of systems (e.g., speech-based tutoring systems~\cite{Pon-Barry2006}).

\begin{table*}
 \centering
 \caption{Codes with highest pointwise mutual information (PMI) with `Uncertainty'.}
 \label{tab:pmi-uncertainty}
 {\footnotesize
	\begin{tabular}{p{.8in}p{2in}p{.25in}p{3in}}
 \toprule
Co-occurring Code & Explanation & PMI & Example \\\hline
Release date & Uncertainty about release date. & 1.153 & I remember a movie about a "super bus", I think it was in the late 70's. \\
Regional origin & Uncertainty about the movie's regional origin. & 0.930 & Can't remember the country of origin but I believe it was Scandinavian. \\
Real person & Uncertaintly about a person associated with the movie. & 0.899 & The girl had a "smiling dimple" that reminded me of Sarah Michelle Gellar but I'm absolutely unsure. \\
Compares music & Uncertainty about movies sountrack. & 0.886 & The lyrics were something like "it's a crazy world" and in the video there are lots of people doing crazy stuff and there is a guy kicking a (fake) dog over a balcony. \\
Singular timeframe & Uncertainty about the movie's timeframe. & 0.836 & I believe it was based in New York and during the 80's. \\
Target audience & Uncertaintly about the target audience. & 0.799 & I believe this is a kids movie I saw on TV in the mid to late nineties. \\
Specific location & Uncertainty about locations in the movie. & 0.758 & All I remember (correctly I hope) is that it was a movie maybe 70's comprised of a number of unrelated scenes all set in Europe. \\
Specific music & Uncertainty about a specific song in the movie. & 0.721 & It was like that song with the rabbit and the three blind mice . \\
Temporal context & Uncertainty about the time when movie was watched. & 0.693 & Here is one I saw long ago \textgreater{}\textgreater 20 years on TV probably a "Creature Double Feature". 
 \\
 \bottomrule
 \end{tabular}}
\end{table*}

\section{Analysing Automatic Retrieval for TOT Requests}
\label{sec:experiment}

In addressing \rqX{1} in the previous section, we  found that searchers in  TOT states used a variety of strategies to express their information need.   In this section, we will be focusing on \rqX{2} and how conventional automatic retrieval systems perform in response to these requests.  Specifically, we are interested in the ability of retrieval systems to support the various tactics adopted by searchers in TOT states.
To this end, we conduct per-code ablation experiments using a standard retrieval system (Okapi BM25~\citep{robertson2009probabilistic}).

\subsection{Methods}
\label{sec:intermediation:methods}
In order to accommodate identification experiments, we need: 
\begin{enumerate*}
\item a collection of \emph{search requests} and the relevant item for each request,
\item a \emph{corpus} where each document is  associated with a unique item,
\item a \emph{retrieval system}, and 
\item an appropriate \emph{evaluation metric}.
\end{enumerate*}
We describe the search requests for our retrieval experiments in Section~\ref{sec:data}.  Because this is an identification task and we only selected TOT requests with a correct answer, each request had exactly one answer (i.e., the correct movie title).

For the corpus, we wanted to assemble a set of indexable items that would be amenable to free text retrieval and the movie identification task.  Long form movie plot descriptions provide a text-rich representation of movies.  We extracted the plot description for each movie in Wikipedia.  We used the \plotdata\footnote{\url{https://github.com/markriedl/WikiPlots}} code to extract plots from a 2019 dump of Wikipedia. We indexed the \plotdata collection with the Indri retrieval system~\citep{strohman2005indri}, removing stopwords using the Indri stopword list and stemming using the Krovetz algorithm~\cite{kstem}.  Each plot description was an average length of 328.9 words after removing stopwords.
Consistent with indexing, the query terms are also stemmed during retrieval.

In order to associate each request with a relevant document in the Wikipedia corpus, we restricted both the requests and the documents to those associated with a unique IMDb identifier.  Several of the correct answers in the `I Remember this Movie' dataset included links to the IMDb page. Additionally, many Wikipedia movie pages include a reference to the movie's IMDb page. Filtering for those Wikipedia entries resulted in 69,132 documents and 339 requests with matches in that corpus.  Each request had exactly one relevant document in the collection.  

As a retrieval method, we adopted Okapi BM25~\citep{robertson2009probabilistic}, allowing a reproducible standard retrieval algorithm. We used a $20\%$ randomly sampled subset of the training set to tune the BM25 parameters.

Finally, we adopted ``success at ten'' as our evaluation metric.  For a given query, this metric is defined as 1 if the correct movie was returned in the top ten positions; 0 otherwise.  This metric reflects the searcher's recall orientation more than mean reciprocal rank, a metric often used for question answering.  

To study the relative usefulness of different phenomena (i.e., qualitative codes) present in TOT requests, we conducted ablation experiments. We adopted the following protocol for this study.  For each code, we first computed the retrieval performance using all the requests that included at least one sentence with that code.  Then, for each of these requests, we computed the retrieval performance after removing all sentences with the code. If performance degrades after removing all sentences with the code, it means that those sentences contributed information that the retrieval algorithm was able to harness to improve results (on average).  Conversely, if performance improves, it means that those sentences included content that degraded retrieval performance. Because some of our codes occur infrequently, we focus only on codes that occurred in $>20\%$ of all TOT requests.

\subsection{Results}
\label{sec:intermediation:results}

Averaging across \emph{all} $339$ requests, we found that only 13.27\% of requests placed the relevant document above the tenth position and 55.16\% retrieved the relevant document above the 1000th position.  We present the results for our ablation experiments in Table \ref{tab:retrieval-annotation}.

In general, sentences descriptive of the content of the movie, when removed, resulted in 9.44\% of requests no longer retrieving the relevant movie in the top ten positions.  However, this impact was not uniform across all content categories.  The most substantial influence came from descriptions of characters, objects, scenes,  locations, and plot summaries; the remaining descriptive types tended to have minimal impact on performance, roughly resulting in negligible changes in success rate.  Sentences with context information (e.g., when and where a movie was seen) similarly had negligible impact on performance.  Finally, sentences expressing uncertainty  \emph{helped} performance.  In other words, we observed an increase in failure of requests after removing sentences where searchers expressed uncertainty.  Neither sentences containing relative comparisons nor those containing social niceties substantially affected retrieval performance.

\begin{table}[]
\caption{Ablation experiments for each annotation label. `Frequency' denotes the percentage of TOT requests containing sentences associated with the code. `All' denotes performance by using the entire TOT request; `ablated' denotes performance by omitting sentences associated with the code. Columns `absolute' and `relative' denote the difference in performance in absolute terms and percent increase/decrease, respectively.  Rows sorted within category by relative difference with respect to the baseline. Larger drops in performance indicate important sentence types.}
\label{tab:retrieval-annotation}
\setlength\tabcolsep{3pt}
\centering
\small{ 
\begin{tabular}{lccccc}\toprule
&&\multicolumn{2}{c}{success@10}&\multicolumn{2}{c}{difference} \\
&	frequency	&	all	&	ablated	&	absolute	&	relative	\\
\hline
Movie (all)	&	100\%	&	0.1327	&	0.0383	&	-0.0944	&	-71.1\%	\\
\tabindent Character	&	98.2\%	&	0.1351	&	0.036	&	-0.0991	&	-73.4\%	\\
\tabindent Object	&	82.3\%	&	0.1434	&	0.0466	&	-0.0968	&	-67.5\%	\\
\tabindent Scene	&	89.7\%	&	0.1382	&	0.0625	&	-0.0757	&	-54.8\%	\\
\tabindent Location type	&	73.7\%	&	0.148	&	0.096	&	-0.052	&	-35.1\%	\\
\tabindent Plot summary	&	61.7\%	&	0.1531	&	0.1148	&	-0.0383	&	-25.0\%	\\
\tabindent Category	&	92.0\%	&	0.1346	&	0.1314	&	-0.0032	&	-2.4\%	\\
\tabindent Genre/tone	&	34.8\%	&	0.0763	&	0.0763	&	0	&	0.0\%	\\
\tabindent Release date	&	43.4\%	&	0.0952	&	0.102	&	0.0068	&	7.1\%	\\
\tabindent Visual style	&	34.2\%	&	0.1552	&	0.1897	&	0.0345	&	22.2\%	\\
\tabindent Language	&	23.9\%	&	0.1111	&	0.1358	&	0.0247	&	22.2\%	\\
Context (all)	&	64.9\%	&	0.1227	&	0.1409	&	0.0182	&	14.8\%	\\
\tabindent Temporal context	&	57.8\%	&	0.1327	&	0.148	&	0.0153	&	11.5\%	\\
\tabindent Physical medium	&	35.7\%	&	0.124	&	0.1488	&	0.0248	&	20.0\%	\\
Other \\
\tabindent Uncertainty	&	88.5\%	&	0.1333	&	0.1067	&	-0.0266	&	-20.0\%	\\
\tabindent Relative comparison	&	20.6\%	&	0.0571	&	0.0571	&	0	&	0.0\%	\\
\tabindent Social	&	51.6\%	&	0.0686	&	0.0743	&	0.0057	&	8.3\%	\\
\bottomrule
\end{tabular}}
\end{table}

\subsection{Discussion}

Our experiments suggest that there is substantial room for improving systems to support TOT requests.  We observed this despite the alignment between our corpus of plot descriptions and the dominant searcher strategy of describing the movie content.  As such, we believe that richer, more granular representations of salient or memorable content should improve retrieval performance.  We note that what is memorable may in fact be different from a straightforward plot description.

Importantly, we observed systematic variation in retrieval performance across different strategies employed by searchers in TOT states.

Because we indexed detailed movie plots, a searcher's description of characters, objects, scenes, and locations allowed for effective retrieval.  However, our results also suggest the importance of content metadata in supporting TOT requests.  Request sentences referring to genre/tone, category, release date, language, and visual style all refer to information about the movie, rather than the content of the movie.  While the lack of effectiveness of types of strategies is likely attributable to missing metadata, we also note that, more generally, metadata descriptions may be too coarse (i.e., applicable to more than one, if not many, movies to result in effectiveness improvements over more precise information specified in a plot description).

Contrary to our expectations, our results demonstrate that sentences expressing uncertainty did \emph{not} degrade performance.  
This indicates that even when the searcher may be unsure, the stated movie description may be accurate and useful for retrieval.
For example, for the sentence ``I saw the movie somewhere around 1986-88 on TV about young (high school) musician boy who played electronic music on a keyboard'', we find term matches with the correct movie plot description that also contains the phrase ``electronic music'', despite uncertainty about the year.  
Similarly, sentences that make relative comparisons can also often include important keywords that match with the target movie plot description.
For example, the sentence ``I don't know if this Laika was the original Laika who travelled to the space in the 50s or the name is in honor of the real astronaut dog'' contains the terms ``space'' and ``dog'' that matches with the target movie plot description.  The inclusion of this sentence in the query formulation on the whole seems to have a positive effect on retrieval---in spite of terms like ``Laika'' appearing in eleven other plot descriptions in our collection.

We were concerned that algorithms may be brittle in the presence of unsupported strategies like metadata or contextual information.  Fortunately, in both cases, we found minimal deterioration in performance after removing sentences with these codes.

Our retrieval experiments were conducted in the context of standard Okapi BM25 model. In future work, it may be interesting to revisit similar research questions in the context of other more sophisticated IR models, such as those that learn latent representations of text~\citep{mitra2018introduction} or operate over structured content~\citep{robertson2004simple, zamani2018neural}.

\section{Conclusion and Future Work}
In this paper, we describe tip of the tongue known-item retrieval, a class of item identification tasks where the searcher has previously experienced or consumed the item but cannot recall an identifier.  Previous research demonstrates that TOT states can be especially frustrating to searchers and, as a result, have led to the creation of community question-answering sites around these needs, covering cultural objects such as movies, music, and books.  Our qualitative coding of a set of TOT requests indicate that searchers employ a variety of information-seeking strategies, including semantic and episodic memories of previous experiences with the item.  Moreover, searchers leverage more sophisticated constructs such as multi-hop reasoning, self-reflective descriptions of previous search attempts, and expressions of uncertainty.  Inspite of the sophistication of these techniques, we found that automatic retrieval was largely unaffected by the presence of these operations.  

We believe that TOT requests reflect an important and open area of information retrieval research. We used movie identification as a case study and several of our observations may or may not exist in domains such as music, books, or other media.  While specific codes we developed may need to be adapted, we suspect that there are more abstract, general behaviors repeated by searchers across other domains. At that, the range of tactics employed during TOT states---perhaps due to frustration---makes this a rich context within which to observe searcher behavior in controlled environments.  This would require the adaptation of TOT elicitation techniques from the cognitive psychology domain to the information retrieval context \cite{Brown:1991aa}.

From an algorithmic perspective, supporting document representations that are both comprehensive (i.e., including detailed descriptions and all metadata) and amenable to more elaborate search strategies will go a long way toward satisfying TOT needs.  As noted in our results, this may require better understanding the distinction between descriptive representations and those biased toward memory-salience.  At the same time, the integration of personal information management and life logging techniques will be necessary for responding to contextual information conveyed by searchers.

\bibliographystyle{ACM-Reference-Format}
\balance{{\bibliography{main}}

\end{document}